\documentclass[epsfig,12pt]{article}
\usepackage{epsfig}
\usepackage{graphicx}

\newcommand{\beq}{\begin{equation}}   
\newcommand{\eeq}{\end{equation}}
\newcommand{\beqn}{\begin{eqnarray}}   
\newcommand{\eeqn}{\end{eqnarray}}

\newcommand{\gsim}{\lower.7ex\hbox{$
\;\stackrel{\textstyle>}{\sim}\;$}}
\newcommand{\lsim}{\lower.7ex\hbox{$
\;\stackrel{\textstyle<}{\sim}\;$}}

\begin{document}

\begin{flushright}
FTPI-MINN-13/33, UMN-TH-3304/13\\
\end{flushright}

\vspace{0.5cm}

\begin{center}
{\Large New and Old about Renormalons:} 

\vspace{3mm}

{\Large In Memoriam Kolya Uraltsev}\footnote{To be published in Nikolai Uraltsev Memorial Volume, (World
Scientific, Singapore, 2014).}

\vspace{2cm}

{\large M. Shifman}

\vspace{0.5cm}
{\em William I. Fine Theoretical Physics Institute, University of Minnesota,
Minneapolis, MN 55455, USA}

\vspace{0.9cm}

{\bf Abstract}

\end{center}

\vspace{0.1cm}

{\small  I summarize what we know of renormalons from the 1970s and 80s: their uses and theoretical status.
It is emphasized that renormalons in QCD are closely related to the Wilsonean operator product expansion 
(OPE) -- a setup 
ideally suited for dealing with the factorially divergent series reflecting infrared dynamics. 
I discuss a breakthrough proposal  due to Uraltsev {\em et al.} to 
use renormalons to evaluate nonperturbative (power) corrections
in the processes without OPE.
Some fresh ideas which were put forward recently are briefly discussed too, with emphasis on  
a possible relationship between resurgence via trans-series and OPE.

 This article is devoted to the memory of my friend Kolya Uraltsev. I should emphasize that these 
 are my personal recollections. Other people who 
 closely knew Kolya may or may not agree with my opinions.
 }

\newpage
\section*{Foreword: \\[2mm] Nikolai (Kolya) Uraltsev}

Kolya, Alexei Anselm's student, was one of the most prominent theorists from the young generation of 
the Gribov Leningrad school.  The heavy quark theory acquired the level of perfection it 
enjoys now to a large extent  due to his works on quantum chromodynamics. In this area, 
there was no higher authority in the world than Kolya.

In science Kolya was a ``slowpoke," in the sense that each new result or new assertion in his field -- 
the theory of heavy quarks -- had to be  critically processed before being accepted (or not). 
Coming across something new Kolya pondered on all sides of this ``something new"  with 
incredible diligence. There was no way any ambiguity could be left after Kolya. He almost 
physically suffered from sloppy works and light-minded authors. Kolya gave no quarter to 
such speakers at seminars or conferences, no matter what regalia they may have possessed. 
With them he was aggressive 
and restless until he had
exposed all  loopholes in the arguments. This ``inconvenient" style -- science above 
everything else -- that Kolya had learned from Gribov, he carried through his life, 
without changing it in the West, where it (the style) is almost extinct due to incompatibility 
with political correctness. Frankly speaking, physicists from the West slightly feared Kolya. 
None of the heavy quark theorists wanted to be ground by ``millstones" in Kolya's mind.

It is ironic that in everyday life Kolya was not only shy, but rather super-shy. 
You can hardly find such shy people nowadays. For him it was a problem to talk to a stranger 
or to respond to the harsh words of an insolent fellow.

Every summer Kolya went on archaeological excavations at the Kola Peninsula. There he met 
his future wife. Well ... this statement is not entirely accurate. Lilya (that's her name) 
at the time worked in Leningrad's Hermitage and also used to go to Kola Peninsula excavations. 
Once she told me: ``Kolya  stared at me for a long time, but did not dare to approach. Then 
I realized that if I do not take matters into my own hands, we will return to Leningrad 
without getting acquainted ..."

When Kolya was thinking about physics, he did not notice anything around him. Once during 
a conference, after a session, we walked out of the conference hall to the street under 
heavy rain. Everybody opened umbrellas right away. Kolya did not react to a change in 
the environment from comfortable to dramatically uncomfortable, and continued the 
discussion as if nothing had happened ... He kept a clean child's soul.

We -- Kolya and I -- published 17 joint works: the first in 1987 and the last in 1998. 
Especially productive was our collaboration during the academic year 1994/95 when Kolya 
spent the whole year with us at the University of Minnesota. Many ideas conceived during 
this year became parts of subsequent research on heavy quark theory. Here I would like to 
single out a particularly exciting insight: the use of renormalons as a tool for revealing  
power terms in the processes without the operator product expansion. 

Kolya could repair with his own hands any damage to any vehicle, including those most 
modern and stuffed with electronics. It was his passionate hobby. In 1996, we spent 
six months together at CERN. For everyday commuting I bought a used Audi, which had 
problems all the time. In the Swiss garages they asked from me exorbitant prices for 
repairs. Kolya coped effortlessly.

In fact, Kolya could fix just about anything, not only cars. In this, like in physics, he 
was inquisitive; he loved the process of learning ``how things work," be it a $B$-meson 
decay or a leak in a boat. 

Striking thoroughness -- that's how I would characterize Kolya's approach to every aspect 
of his life and work. During his 30-year career in theoretical physics Kolya closely and 
productively interacted with many colleagues on three continents,  St. Petersburg, CERN, 
Technion, Milan, Orsay, FTPI, University of Notre Dame, and University of Siegen.  I 
am sure that's how they will remember him -- a deep thinker and a reliable friend.

Death is always untimely. When Kolya's heart stopped on February 13, 2013 he was only 
54 years old, full of plans for the future both in science and life.
Even now, six months after his tragic death, it is not easy for me to write this 
{\em in memoriam} article in a 
logically-ordered manner.  Apparently, I will have to settle for less.

\section{Introduction}

One can say that Kolya burst onto heavy quark theory like a meteor. Our first (occasional) 
scientific encounter occurred in 1986 
\cite{khoze}. Shortly after, our paths departed: he delved in the problem of $CP$ violation 
for six long years \cite{bigi1,bigi2}, while I returned to nonperturbative supersymmetry \cite{mashi}. 
I was still heavily involved in this topic when Kolya appeared in our 
Institute\footnote{William I. Fine Theoretical Physics Institute, University of Minnesota.} 
in 1992 full of enthusiasm with regards to a consistent theory of  $1/m_Q$ expansion
in heavy quarks based on the operator product expansion (OPE). Elements of this theory 
already existed \cite{A,B,IW}. However, they represented general guidelines rather than 
a theoretical construction worked out in detail. Applications were rather
scarce. Kolya's enthusiasm was contagious, and shortly after both, Arkady Vainshtein 
and myself,  got fully involved.  
One of the most elegant results established by Kolya and collaborators
\cite{Vai}  (see also \cite{Gri}) was  the 
absence of the $1/m_Q$ correction in the total inclusive decay widths of 
the  heavy-flavor hadrons.
This theorem (sometimes referred to as the CGG -- BUV theorem)    made its way into textbooks,
let alone its practical importance for precision determination of 
$V_{cb}$ and $V_{ub}$ from data.  

In twenty years that elapsed after 1992 Kolya managed to make definitive contributions to 
many topics from the heavy quark theory. It is fair to say that he left no stone unturned. 
His imprint is seen everywhere. Needless to say, I will be unable to cover all these topics. 
Instead, I will focus on one particular topic -- determination of the power of $1/m_Q$ 
(or $1/Q$) terms from renormalons in the processes without OPE -- in which Kolya was a 
trailblazer.\footnote{
The reader unfamiliar with the range of questions associated with OPE, $1/m_Q$ corrections 
in heavy quark theory, renormalons and all that is advised to turn to reviews 
\cite{ShifHQE,beneke,ManWis,NSVZo,snap}.}

\section{Pioneers}

Two papers, \cite{polemass} and \cite{bb},   
which appeared on ArXiv on the same day, were the first to
suggest the usage of renormalons for indication of the power of nonperturbative corrections 
(i.e. $1/Q$ or $1/m_Q$, or squares, cubes etc. of the above parameters)
in  the processes {\em without} OPE.

The next relevant paper was \cite{web}, where the idea was first applied  to hadronic 
event shapes. 

Before explaining how this works, I will have to remind you what renormalon is. To this 
end I will have to start from the
factorial divergences of the perturbative series.

\section{Dyson argument and factorial divergences}
\label{da}

Sixty two years ago Freeman Dyson completed his famous paper entitled
``Divergences of Perturbation Theory in Quantum Electrodynamics" \cite{dyson} 
(reprinted in \cite{LOBPT}).
He argued that the series in $e^2$ in QED could not be  convergent due to the fact 
that analytic continuation
to negative $e^2$ produced a theory with unstable vacuum. This became known as the 
Dyson argument. Shortly after, Thirring evaluated \cite{thir}  the 
number of diagrams in $\lambda \phi^3$ field theory in high orders and  came to the conclusion
that the perturbative series in this theory is factorially divergent. In 1977 various field 
theories, including $\lambda \phi^4$,  were thoroughly
studied by Lipatov \cite{lip} who came to the same general 
conclusion:  the  perturbative series are asymptotic and characterized by the factorial divergence 
of the  form
\beq
Z = \sum_k C_k\,\alpha^k\, \,k^{b-1}\, A^{-k} k!\,.
\label{fd}
\eeq
This is reviewed in some detail e.g. in \cite{LOBPT}. The notation 
in Eq. (\ref{fd}) is as follows: $\alpha$ is the expansion parameter,\footnote{In QED it is 
customary to define
$\alpha \equiv e^2/(4\pi)$. The asymptotic divergence of the coefficients in  QED is 
somewhat more contrived \cite{Bogomolny:1978ft}  than in (\ref{fd}) 
due to the fact that the QED loops are due to fermions.}
$k$ is the number of loops, $C_k$'s are numerical coefficients of order one,
and $b$ and  $A$ are numbers.

Arkady Vainshtein was the first to point out \cite{ava} (see \cite{avae}) that the factorial 
divergence in (\ref{fd}) is in one-to-one correspondence with the probability of the 
under-the-barrier penetration (vacuum
instability in field theory language) for unphysical -- negative -- values of the 
expansion parameter. Ten years later this relation was rediscovered by Bender and Wu \cite{bw} 
in the quartic anharmonic oscillator or, which is the same, in $\lambda\phi^4$ theory.

The factorial divergence of the perturbative series discussed in \cite{dyson,LOBPT,ava,avae,bw} 
 can be traced back to the
factorially large number of multiloop Feynman diagrams (i.e. $k\gg 1$).

Renormalons which we will focus later have nothing to do with this mechanism. As 
was noted in \cite{renormalon}, there exists a class of isolated graphs, in which each diagram 
grows factorially as we increase the number of loops. It is these graphs that are called renormalons. 
The theoretical feature responsible for the 
 renormalon factorial  divergence (\ref{fd})
is the logarithmic running of the effective coupling constant.

\section{Borel summability}
\label{bors}

Instead of the {\em asymptotic series} (\ref{fd}) let us introduce the Borel transform
\beq
B_Z(\alpha) = \sum_k C_k\,\alpha^k\, \,k^{b-1}\, A^{-k}\,.
\label{bt}
\eeq
In Eq. (\ref{bt}) the $k$-th term of expansion (\ref{fd}) is divided by $k!$, which implies, 
in turn, that the singularity of 
$B_Z(\alpha) $ closest to the origin in the $\alpha$ plain is at distance $A$ from the origin.  
Thus, the sum (\ref{bt}) is convergent. 

   Mathematicians would say that the function defined by (\ref{bt}) is obtained from (\ref{fd}) 
   by the inverse Laplace transformation.

It is quite obvious that one can recover the original function $Z$ performing the following 
integral transformation
(the Laplace transformation): 
\beq
Z(\alpha) = \int_0^\infty \, dt\, e^{-t} B_Z(\alpha\,t)\,,
\label{bti}
\eeq
see e.g. \cite{JZJ}, Sect. 37.3. The integral representation (\ref{bti}) is well-defined 
provided that $B_Z(\alpha)$ has no singularities
on the real positive semi-axis in the complex $\alpha$ plane. This is the case if the asymptotic 
series (\ref{fd}) is sign-alternating, $C_k\sim (-1)^k$,  (and then so is
(\ref{bt})). If $B_Z(\alpha)$ has singularities on the real positive semi-axis (as is the 
case if the  coefficients
$C_k$ are all positive, or all negative), then the integral (\ref{bti}) becomes ambiguous. 
The ambiguity is of the order of
$e^{-A/\alpha}$.
One cannot  resolve this ambiguity on the basis of purely mathematical arguments. 
More information is needed, which can be provided only by underlying physics. 

In problems at weak coupling additional physical information can be obtained by quasiclassical methods. 
Indeed, at weak coupling deviations from perturbation theory are due to classical solutions 
with nonvanishing action, such as instantons or instanton-antiinstanton (IA) pairs. 
Say, in the quantal problem of the double-well potential, the contribution of the  
instanton-antiinstanton pair is ambiguous {\em per se}. However, one can combine
(\ref{bti}) with the latter in such a way, that in the final answer these two ambiguities 
cancel, giving rise to a well-defined expression \cite{bo,bo1}. The next ambiguity occurs at the 
level of two instanton-antiinstanton pairs. It is canceled against the ambiguity in perturbation theory 
in the sector of a single instanton-antiinstanton pair plus a subleasing singularity \cite{au}
in (\ref{bti}). The process of cancellation of ambiguities is repeated {\em ad infinitum}. 
Continuing this procedure one arrives at the so-called trans-series combining perturbative and 
quasiclassical nonperturbative expansion at weak coupling. In a slightly simplified form the 
resurgence and trans-series
can be expressed by the formula
\beq
Z(\alpha ) = \sum_{k=0}^\infty\,\left\{c_{0,k} + c_{1,k} \alpha + c_{2,k} \alpha^2+c_{3,k} \alpha^3 + ...
\right\}e^{-kA/\alpha}\,,
\eeq
where for each given $k$ the coefficients $c_{n,k}$ are factorially divergent in $n$, and 
the sum in $n$ in the braces (for each given 
$k$) is regularized in a well-prescribed manner. I will say a few words on the nature of 
the $k$ series later.

In quantum mechanics the construction of the trans-series was explored in
\cite{bo,bo1,au,bo2,bo3,Aniceto}. Recently  a progress along these lines was achieved in field 
theory too \cite{bo4,bo5}.

To make sure that a field-theoretical model under consideration is weakly coupled, it 
was analyzed \cite{bo4,bo5} in cylindrical geometry $R_1\times S_1 (  r  )$, with a 
compactified dimension 
of a very small size $r$. Then, in much the same way as in the above quantal problem, 
it proved to be possible to identify  quasiclassical field configurations responsible 
for nonperturbative contributions \cite{u1,u2}, to be combined with the Borel-resummed perturbative series.

It is quite plausible that in weakly coupled field theories a complete resurgence 
can be achieved along these lines,
and at least some quantities are representable in the form of trans-series
combining Borel-resummed perturbation theory with a (infinite) set of nonperturbative effects derivable
from quasiclassical considerations. What remains to be seen is whether this program works in
a more general setting of any weakly coupled field theory, for instance, in fully Higgsed Yang-Mills 
theory, and if yes,
in which particular way.
At the moment
the idea of matching the factorial divergence to quasiclassical field configurations in 
fully Higgsed Yang-Mills theories is barely explored.\footnote{Some hypotheses 
are discussed in Sect. \ref{riwcp}.} 
 
If this idea survives in a more general formulation, the next intriguing question is obvious:
whether or not a connection to strong coupling regime can be revealed. 
Note that at weak coupling continuous symmetries such as the chiral symmetry cannot be spontaneously broken.
Therefore, a parallel between resurgence via trans-series in quantum mechanics on the one hand
and OPE in QCD and similar theories on the other, which of course comes to one's mind, cannot be complete.  

\section{The first source of factorial divergence}
\label{ibafd}

In quantum mechanics the coupling constant is fixed. In Yang-Mills field theory (e.g. QCD) 
the very notion of the smallness
of the coupling constant is meaningless, since the coupling constant depends on scale; it runs and 
becomes strong at
momenta of the order of dynamical scale $\Lambda$. At such momenta
dynamics are by no means exhausted by perturbation theory
and quasiclassical nonperturbative effects. 
In fact, in the infrared domain, at strong coupling, both cannot even be 
consistently defined.
Below we will discuss what can be done under the circumstances. 

For a short while, let us close our eyes at this feature pretending that somehow the blow 
off of $\alpha_s$ in the infrared (IR) domain is not essential. This neglect will be corrected shortly.
In Yang-Mills theory one can identify at least two sources for the factorial 
divergence of the perturbative series.
First, the number of various Feynman graphs with $n$ loops grows as $n!$. 
This feature (similar to that one encounters in
quantum mechanics)
was known already to the explorers of QED from the times of the Dyson argument, see Sect. \ref{da}. 
As a result, even if each graph is of the order of unity in appropriate units, 
the contribution of the set of the $n$-loop graphs will be of the order of $n!\,\alpha^n$. 
At $n\sim 1/\alpha \gg 1$ multiloop graphs are typically represented by soft 
fields which can be viewed as quasiclassical field configurations, for instance, instantons. 
Instanton contributions to correlation functions are 
$\sim \exp\left(-\frac{2\pi}{\alpha}\right)$. 

If this were the only source, the problem could be eliminated in an elegant way,
which can be traced back to 't Hooft's observation \cite{thooft} that in the limit
\beq
N\to \infty \,,\quad N\alpha \,\,{\rm fixed}\,,
\label{z1}
\eeq 
where $N$ is the number of colors, only planar diagrams survive. The limit (\ref{z1}) is 
referred to as the 't Hooft limit.
Three years after 't Hooft's original work a remarkable theorem was proved \cite{koplik}: 
the number of planar diagrams 
with $n$ loops $\nu (n)$ does {\em not} grow with $n$ factorially, rather
\beq
\nu (n) \sim C^n\,,\quad n\gg 1\,,
\eeq
where $C$ is a numerical constant. In one-to-one correspondence with this fact is the 
vanishing of the instanton contribution at $N\to\infty$. Indeed, at weak coupling in the 't Hooft limit
$$
\frac{2\pi}{\alpha}\sim {\rm const}\, N\,,
$$ 
and the instanton contribution is exponentially suppressed.

One can identify another source of the factorial divergence -- unique diagrams of a special 
type present in 
Yang-Mills  which produce
$n!$ not because there are many of them, but because a single graph with $n$ loops is factorially large. 
As was mentioned previously, such diagrams are called renormalons \cite{renormalon,renormalonr}. 
In the subsequent section
we will consider them in more detail.

\section{Renormalons}
\label{reno}

\begin{figure}[t]
   \vspace{-3.3cm}
 \epsfysize=30cm
   \epsfxsize=20cm
   \centerline{\epsffile{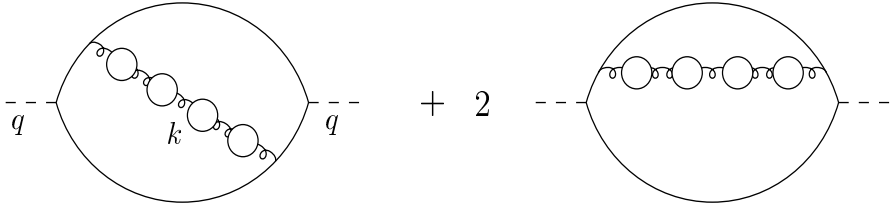}}
\vspace*{-22.8cm}
\caption{\small Bubble diagrams for the 
Adler function consists of all diagrams with any number of fermion 
loops inserted into a single gluon line. 
Then we replace $\beta_0^{f}$, the fermion contribution to the first coefficient of the 
$\beta$ function by
the full $\beta_0$. This is a convenient computational device.
\label{fig1}}
\end{figure}

Both, ultraviolet (UV)
 and IR renormalons can be seen in the bubble diagram depicted in Fig. \ref{fig1}, where the dashed line 
represents an external (vector) fermion current, the solid lines show fermion propagation while the curvy 
lines stand for gluons.
Consider the correlation functions of two vector currents  of massless quarks
\begin{eqnarray}
\Pi_{\mu\nu} (q) &=&i\, \int\,d^4x\,e^{-i q x}\,\left\langle T\,
\left[ j_\mu(x) j_\nu(0)\right]\rule{0mm}{4mm}\right\rangle 
= \left(q_\mu q_\nu-q^2 g_{\mu\nu}\right)\,\Pi(Q^2)\,,
\nonumber\\[2mm]
j_\mu &=& \bar\psi\gamma_\mu\psi\,,
\label{currentcorr}
\end{eqnarray}
where $\psi$ is the quark field; we assume the number of flavors to be $N_f$, and denote
 \beq
 Q^2=-q^2\,,
 \label{eucl}
 \eeq
 so that in the Euclidean domain $Q^2$ is positive.
  The number of colors $N_c=3$. It is convenient to analyze the Adler function defined as
\begin{equation}
\label{adlerdef}
D(Q^2)=- 4 \pi^2\,Q^2\,\frac{d\Pi(Q^2)}{dQ^2}
\end{equation}
and normalized to unity in the leading order.
The bubble diagrams with the fermion loop insertions are gauge-invariant {\em per se}.
Needless to say, in the given order there are many other graphs, but we will focus on
those presented in Fig. \ref{fig1}, which will be sufficient for identification of renormalons.

The gauge coupling runs, and we must specify which particular coupling constant is used
in the expansion. The Adler function, being expressed in terms of $\alpha_s (Q)$, is finite. 
It seems obvious that the external momentum
$Q^2$ sets the scale of all virtual momenta in loops, and we should use $\alpha_s (Q)$. 
Is it indeed the case?
 
The answer to the above question is negative. In high orders of perturbation theory there appears 
an additional parameter $n$,
the order of perturbation theory, which changes the naive estimate $k\sim Q$, see Fig. \ref{fig1}.
To see that this is the case, let us have a closer look at Fig. \ref{fig1} before integrating over $k$. 
The exact result for fixed $k^2$ was found by Neubert \cite{MNe}. However, we will not need it since 
for our illustrative purposes it is sufficient
to use a simplified interpolating expression \cite{vzak} collecting all fermion bubble 
insertions\footnote{The fermion bubbles in Fig. \ref{fig1}
produce only the fermion contribution to $\alpha_s(k^2)$ usually denoted by $\beta_0^f$. However, 
then we can replace
$\beta_0^f$ by the full $\beta_0$. Note that adding the gluon and ghost {\em bubbles} is not 
sufficient (in particular, one would  get a gauge noninvariant expression). The replacement 
$\beta_0^f\to \beta_0$ incorporates some additional contributions. Note that $\beta_0^f$ 
and $\beta_0$ have opposite signs -- a crucial feature
as we will see below.}
in the gluon propagator: $0,\,1,\,2$ and so on,
\beq
D= C \times Q^2 \int
dk^2\frac{k^2\alpha_s(k^2)}{(k^2+Q^2)^3} \,,
\label{renorm}
\eeq
which coincides with the exact expression \cite{MNe} in the limits
$k^2\ll Q^2$ and $k^2\gg Q^2$, up to minor irrelevant details.
The coefficient $C$ in Eq. (\ref{renorm}) is a numerical constant and $\alpha_s(k^2)$
is the running gauge coupling constant,\beq 
\alpha_s (k^2) = \frac{\alpha_s (Q^2)}{1-\frac{\beta_0\alpha_s (Q^2)\rule{0mm}{4mm}}{4\pi} 
\ln (Q^2/k^2)} \,.
\label{eight}
\eeq
The definition of the coefficients in the $\beta$ function is given in Appendix.

  Now, let us examine the Adler function (\ref{renorm}) paying special attention
  to the logarithmic dependence in (\ref{eight}), a crucial feature of QCD. 
 We will first focus on the IR domain. Omitting the overall constant $C$, inessential for our
 purposes, we obtain 
  \beq
  D (Q^2)= \frac{1}{Q^4} \,\alpha_s \, \sum_{n=0}^\infty \left(\frac{\beta_0\alpha_s}{4\pi} 
  \right)^n  \int\,dk^2 \,k^2 \left(\ln \frac{Q^2}{k^2}
  \right)^n\,,\qquad
  \alpha_s \equiv \alpha_s (Q^2)
  \label{nine}
  \eeq
  which can be rewritten as
   \beq
  D (Q^2)=  \frac{\alpha_s}{2}\,\sum_{n=0}^\infty  \left(\frac{\beta_0\alpha_s}{8\pi} 
  \right)^n  \int\,dy \,
    y^n\, e^{-y}\,,\qquad
 y = 2 \ln \frac{Q^2}{k^2}\,.
  \label{nineppp}
  \eeq
The $y$ integration in Eq. (\ref{nineppp}) represents all diagrams of the type depicted
in  Fig.~\ref{fig1} after integration over the loop momentum $k$
of the ``large" fermion loop (and the angles of the gluon momentum). 

\begin{figure}[t]
   \epsfxsize=10cm
   \centerline{\epsffile{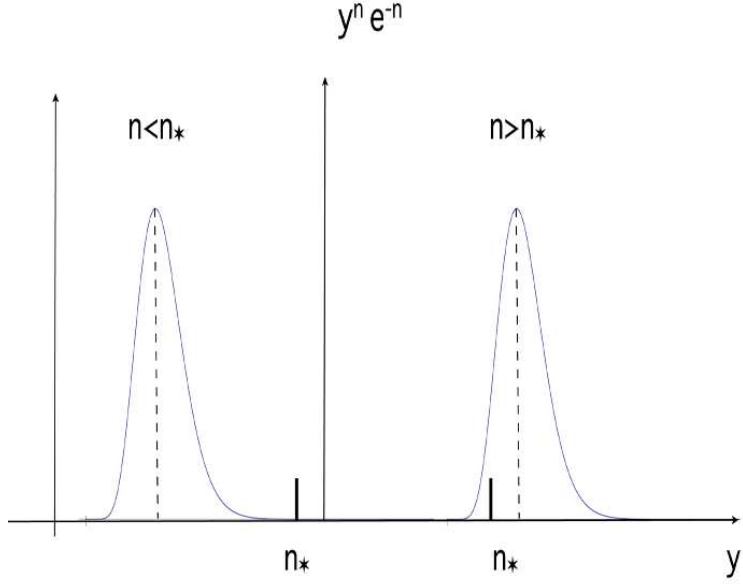}}
\caption{\small The plot of the integrand in Eq. (\ref{nine}) for two values of $n$, ``small" 
and ``large." A sharp peak at $y\sim n$
saturates the integral. In the left plot $n<n_*= 2\ln(Q^2/\Lambda^2)$ and the forbidden 
domain $k^2\sim\Lambda^2$ does not contribute to the factorial factor.  In the right plot $n>n_*$. The $y$ integration has to be cut off at $y=n_*$,
which tempers the factorial growth.
}
\label{fig2}
\end{figure}

The $y$ integral from zero to infinity is $n!$. A characteristic value of $k^2$
saturating the integral is
\beq
y\sim n   \,\,\,{\rm or}\,\,\,   k^2 \sim Q^2\,\exp \left(-\frac{n}{2}\right)  \,.
\label{ten}
\eeq
Thus, if $Q^2$ is fixed and $n$ is sufficiently large, the factorial divergence of 
the coefficients in
 (\ref{nine}) is indeed due to the infrared behavior
in the integral (\ref{renorm}). For what follows let us note that
if at small $k^2\sim \Lambda^2$ the diagram in Fig.~\ref{fig1} ceases to properly 
represent non-Abelian dynamics (which {\em is} the case in QCD due to strong coupling in 
the IR), then the integral must be cut off from below at $k^2 =\Lambda^2$, or at $y=n_*$ 
at large $y$. Here for each given $Q^2$
\beq
n_* = 2\ln\frac{Q^2}{\Lambda^2}\,.
\label{r2}
\eeq
The summation of factorially divergent terms in the formula
\beq
 D (Q^2)=  \frac{\alpha_s}{2}\,\sum_{n=0}^\infty  \left(\frac{\beta_0\alpha_s}{8\pi} \right)^n  n! 
 \label{r1}
\eeq
ceases to be valid at $n=n_*$.
At $n>n_*$ the factorial growth is suppressed, see Fig. \ref{fig2},
and must be truncated,
\beq
 D (Q^2)\to   \frac{\alpha_s}{2}\,\sum_{n=0}^{n_*} \left(\frac{\beta_0\alpha_s}{8\pi} \right)^n  n! \,.
 \label{r2}
\eeq
Note that $n_*$ is also the critical value of
the asymptotic series (\ref{r1}), i.e. the value at which the accuracy of approximation is the best.
At $n=n_*$ the asymptotic series (\ref{r1}) achieves the highest accuracy. Truncation 
at $n=n_*$ ensures  the deviation from the exact result to be 
$\exp\left(-\frac{8\pi}{\beta_0\alpha_s}\right) \sim \Lambda^4
/Q^4$, the same as the infrared sensitivity to the domain $k^2\sim \Lambda^2$.

All terms in (\ref{r1}) have the same sign, which means that the asymptotic 
series {\em per se} is not Borel-summable.

Now let us briefly consider the large $k^2$ domain in (\ref{renorm}).
At large $k^2$
\beq
  D (Q^2)= {Q^2} \,\alpha_s \, \sum_{n=0}^\infty \left(\frac{\beta_0\alpha_s}{4\pi} \right)^n  (-1)^n \int\,dk^2 \frac{1}{(k^2)^2} \left(\ln \frac{k^2}{Q^2}
  \right)^n\,.
  \label{ninep}
  \eeq
Introducing
\beq
\tilde y = \ln \frac{k^2}{Q^2}
\eeq
we arrive at 
\beq
  D (Q^2)= \alpha_s \, \sum_{n=0}^\infty \left(\frac{\beta_0\alpha_s}{4\pi} \right)^n  (-1)^n \int\,d\tilde y
  {\tilde y}^n \, e^{-{\tilde y}}=  \alpha_s \, \sum_{n=0}^\infty \left(\frac{\beta_0\alpha_s}{4\pi} 
  \right)^n  (-1)^n\,n!\,.
  \label{ninepp}
  \eeq
  This series is sign-alternating and, hence, is Borel-summable. The characteristic value 
  of $\tilde y$ saturating the integral
  is $\tilde y \sim n$ implying $k^2\sim Q^2 \,e^n$. Thus, at large $n$ we deal with 
  large $k^2$ which explains
  why this contribution is referred to as the ultraviolet renormalon. It is well-defined {\em per se}.
  The best possible accuracy one can achieve with Eq. (\ref{ninepp}) compared to the exact result
  of the Borel transformation (\ref{bti}) is $\exp\left(-\frac{ 4\pi }{\beta_0\alpha_s} 
  \right)\sim \frac{\Lambda^2}{Q^2}$. 
 I will not touch UV renormalons in what follows.~Note that the singularities in the 
 Borel-transform for the IR and UV renormalons have different separations 
  from the origin.\footnote{As was mentioned above, Eq. (\ref{renorm}) is simplified. 
  Working with the 
  exact formula \cite{MNe}  we would have obtained for the UV renormalon contribution 
  in $B_D$ an expression  
  that contains a single pole as well as a double pole at $\left(\alpha_{s}\right)_* = -  \frac{4\pi}{\beta_0}$.
  In addition to the term in the second line in (\ref{m1}) we would get another term $\sim 
  \left(1+ \frac{\beta_0\alpha_s}{4\pi}\right)^{-2}$, see e.g. \cite{beneke}. 
   No qualitative
  changes occur due to  the
  presence of the double pole. 
I will not go into details,
  since the UV renormalons are mentioned only for completeness and will not be pursued further.  }
  Namely,
  \beq
  B_D (\alpha_s) \sim\left\{\begin{array}{l}
  \left(1-\frac{\beta_0\alpha_s}{8\pi}\right)^{-1}\,, \quad {\rm IR}\,, \\[3mm]
  \left(1+ \frac{\beta_0\alpha_s}{4\pi}\right)^{-1}\,, \quad {\rm UV}\,.
  \end{array}
  \right.
  \label{m1}
  \eeq
  
\begin{figure}[t]
   \epsfxsize=10cm
   \centerline{\epsffile{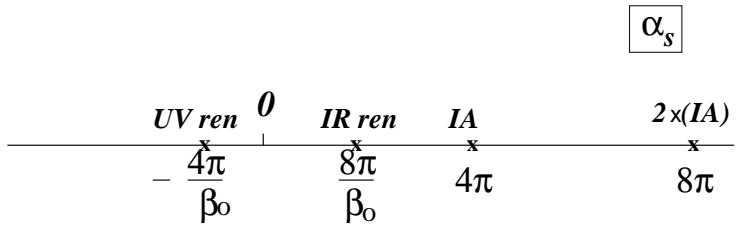}}
\caption{\small Singularities in the Borel plane.
}
\label{fig3}
\end{figure}

  The positions of the singularities in the $\alpha_s$ plane are
  \beq
\left(\alpha_{s}\right)_* = \left\{\begin{array}{l}
 \,\,\,  \frac{8\pi}{\beta_0}\,, \quad {\rm IR}\,, \\[3mm]
-  \frac{4\pi}{\beta_0}\,, \quad {\rm UV}\,.
  \end{array}
  \right.
  \label{m2}
  \eeq
They are depicted in Fig. \ref{fig3} along with the singularities due to instantons.

\section{Operator product expansion}

I remember that after the first seminar on the SVZ sum rules \cite{SVZ} in 1978 Eugene 
Bogomol'nyi used to ask me each time we met:
``Look, how can you speak of power corrections in the two-point functions at large $Q^2$ 
when even the perturbative expansion (i.e. the expansion in $1/\ln(Q^2/\Lambda^2)$) is 
not well defined? Isn't it an `excessive force'  abusive approach?" 

At that time Wilson's wisdom only started to conquer the high-energy physics community. 
My understanding in 1978 missed some nuances  too. In modern terminology what  we mostly
used in \cite{SVZ}  is now known  as the ``practical"  operator product expansion, its 
simplified version (see e.g.
\cite{snap}, Sect. 6, which is highly recommended to the reader unfamiliar with
nuances of OPE in QCD). Although it is sufficient  in solving problems arising in various 
applications and relatively easy to implement, in the present article in which I am after 
conceptual aspects,  I have to be careful with formulations.
Only then I will be able to answer the above Bogomoln'yi question in a positive way, namely: 

``Consistent use of Wilson's OPE makes
everything well-defined  at the conceptual level. Technical implementation may not always 
be straightforward, however." 
 
The operator product expansion (OPE) in asymptotically free theories is a {\em book-keeping device}
separating short-distance (weak-coupling) contributions from
those coming from large distances (strong coupling domain). To this end one introduces an auxiliary
 separation scale $\mu$.
OPE is applicable whenever one deals with problems that can be formulated in the Euclidean 
space-time and in which
one can regulate typical Euclidean distances by a varying large external momentum $Q$ 
(or $m_Q$ in the heavy quark problems).  
Wilson's OPE is meaningful if one can choose $ \mu\ll Q$ (or $\mu\ll m_Q$), but $\mu\gg\Lambda$.

Note that
a practical version that draws a divide in OPE between perturbation theory and 
nonperturbative effects 
is a simplification which may or may not be approximately valid, depending on the theory 
under consideration.
The correct divide is between short- and large-distance contributions. 
As a book-keeping device of this type it cannot fail \cite{NSVZo}, provided no arithmetic 
mistake is made {\em en route}.
   
The idea of factorization of short and large distances, the central point of 
OPE,  dates back to classical  Wilson's work  \cite{wilson} (see also \cite{WKog}) where it was 
put forward in connection with theories of strong interaction with conformal invariance
at short distances.  Shortly after, Wilson formulated a very general 
procedure of the renormalization-group flow (e.g. \cite{WKog}) which became 
known as the Wilsonean renormalization group. Wilson's formulation makes 
no reference to perturbation theory, it applies both to strongly and 
weakly coupled theories. The focus of Wilson's work was on statistical 
physics, where the program is also known as the block-spin approach.
 Starting from the microscopic degrees of freedom at the shortest 
distances $a$, one ``roughens" them, step by step, by constructing a 
sequence of effective (composite) degrees of freedom at distances $2a$, 
$4a$, $8a$, and so on. At each given step $i$ one constructs an effective 
Hamiltonian,
which fully accounts for dynamics at distances shorter than $a_i$
in the coefficient functions.  

Surprisingly, in high-energy physics of the 1970s the framework of OPE was narrowed 
down to a very limited setting. On the theoretical side,
it was discussed almost exclusively  in perturbation theory, as is seen, 
for instance, from Refs.
\cite{pope}. On the practical side, its  applications were mostly narrowed down to 
deep inelastic scattering, where it was customary to work 
in  the leading-twist approximation. 

The general Wilson construction was adapted to QCD, for the 
 systematic inclusion of power-suppressed
effects, in \cite{SVZ,NSVZo}.  Vacuum expectation value of the gluon density operator
and other vacuum condensates were introduced for the first time, 
which allowed one to analyze
a large number of vacuum two- and three-point functions, with quite nontrivial results.
A consistent Wilsonean  approach requires an auxiliary 
normalization point $\mu$ which plays the role of 
a ``regulating" parameter separating hard contributions included in the 
coefficient functions and soft contributions residing in local operators
occurring in the expansion. The degree of locality is regulated by the same parameter
$\mu$.

 Prevalent in the 1970s was a misconception that
the OPE coefficients are determined exclusively by perturbation theory
while the matrix elements of the operators involved are purely 
nonperturbative. Attempts to separate
perturbation theory from ``purely nonperturbative" condensates
gave rise to inconsistencies (see e.g.
\cite{David}; I will return to this paper later) which questioned the very possibility of using the 
OPE-based methods in QCD.  

In the heavy quark theory, in which Kolya's contribution was instrumental,
OPE acquired a new life constituting the basis
of the heavy quark mass expansions (for a review see \cite{ShifHQE}). 
In this range of questions one deals with expectation values of various operators over the
heavy quark meson or baryon states, rather than  vacuum expectation values. 
The overall ideology does not change, however.

 The OPE formalism  provides a natural framework for the discussion of IR renormalons and 
 how they should be treated in theories with strong coupling regime.

\section{An illustrative example}
\label{illu}

Despite the conceptual simplicity of OPE, it continues to be questioned in the literature, 
in particular, 
in connection with renormalons in strongly coupled theories. The statement which I would 
like to illustrate
in this section is: if one introduces the boundary point $\mu$ (unavoidable in non-conformal 
field theories) and abandons the idea of separation along the line ``perturbative vs. 
nonperturbative," all would-be inconsistencies disappear, and so does the problem of renormalons. 

Following \cite{NSVZo} I will consider 
here a relatively simple example of a two-dimensional model -- the so-called $O(N)$ model -- 
which has both, asymptotic freedom and renormalons, and at the same time is exactly solvable at large $N$. Classically excitations in this model are massless. A mass gap is generated at the quantum (nonperturbative) level. 
 This example in the given context was suggested long ago in \cite{Davidp}. In this paper
OPE (in its ``practical" version) 
was found to be perfectly consistent with the exact solution in the leading in $1/N$ approximation. 

However, the subsequent exploration of composite operators \cite{David} questioned the existence 
of consistently defined composite operators in OPE  at the level of the first subleading correction 
(of the relative order of $1/N$). Now,
I will demonstrate how inconsistencies are eliminated once $\mu$ is explicitly introduced. 

The Lagrangian of the model has the form \cite{on} (for a review see \cite{Novikov})
\beq
{\mathcal L} =\frac{N}{2\lambda} \left(\partial_\mu S^a\right) \left(\partial^\mu S^a\right)\,,
\qquad \vec S^{\,2} = 1\,,
\eeq
where $\vec S = \{ S^1, S^2, ..., S^N\} $ is an $N$-component real (iso)vector field, and $\lambda $ 
is the 't Hooft coupling,
which stays fixed in the limit $N\to\infty$. The O$(N)$
symmetry of this Lagrangian is evident. This model is asymptotically free \cite{Novikov},  
in much the same way as Yang-Mills theory,
\beq
\lambda (p) = \frac{2\pi}{\ln \frac{p}{m}}\,, \,\,\, {\rm or}\,\,\, m=p\,\exp\left[-\frac{2\pi}{\lambda (p)}
\right],
\label{22}
\eeq
where $m$ is a dynamically generated mass gap. In perturbation theory the O$(N)$
symmetry is spontaneously broken implying $N-1$ Goldstone modes. The  O$(N)$ symmetry is 
restored in the exact solution, in full accord with the Coleman theorem
\cite{Coleman}. The vacuum condensates of the
type $\langle  \left(\partial_\mu S^a\right) \left(\partial^\mu S^a\right)\rangle\neq 0$ develop. 
To the leading order in $N$ (for details see e.g. \cite{Novikov})
\beq
\left\langle\left[\left(\partial_\mu S^a\right)^2 \right]^k\right\rangle =m^{2k}\,,\quad k=1,2,...
\label{23}
\eeq
In this order the above matrix elements scale as $N^0$ and factorize.  To order $O(N^0$) 
each of them is $\mu$
independent because in this order the anomalous dimension of the operator 
$\left(\partial_\mu S^a\right)^2$ vanishes.
Needless to say, there are
nonfactorizable corrections scaling as $1/N$. For what follows it is convenient to 
introduce a special notation for the operator 
\beq
  \left(\partial_\mu S^a\right)^2 \equiv   \alpha\,.
\eeq
The  operator basis  in OPE  to the order $O(N^0)$ consists of
the composite operators of the type $\alpha^n$. In the subleading orders operators with an 
entangled index structure appear, but we do not have to consider them here. 

To discuss OPE let us consider the two-point function
\beqn
P (q^2) &=&
i\int \, d^2x\, e^{iqx} \left\langle T\left\{ j_s (x) \, j_s (0)\right\}\right\rangle\,,\nonumber\\[2mm]
j_s&=& \sqrt{N} \left(\partial_\mu S^a\right)^2
\label{24}
\eeqn
at large (Euclidean) values of $q^2$ (i.e. $q^2$ negative and $Q^2\equiv - q^2$ positive). 

The general OPE formula for the
two-point function (\ref{24}) (at large Euclidean $Q^2$) has the form
\beq
P(Q^2) =  {c_0(Q^2,\mu^2)}\,{Q^2} \,\,{\rm \bf I}+ {c_1(Q^2,\mu^2)}\,\alpha(\mu) +
\frac{c_2(Q^2,\mu^2)}{Q^2}\,[\alpha(\mu)]^2+ ...
\label{34}
\eeq
where $c_i$ are the coefficient functions.

With our normalization $P(Q^2) \sim N^0$ in the leading order in $N$. All coefficient 
functions and expectations values
scale in the same way, as $N^0$, with subleading $1/N$ corrections.
Moreover, to the leading  order the OPE coefficients  in (\ref{24}) are $\mu$ independent, 
with no factorial divergences. As a result, at $N = \infty$ one can close one's eyes on 
subtleties and adhere to the  simplified formula according to which the coefficient functions are 
determined exclusively by perturbation theory,
and (large-distance) vacuum condensates exclusively by nonperturbative 
effects.\footnote{This exceptional situation
specific to the $O(N)$ model has no parallel in QCD.} This simplified formula is 
self-consistent \cite{Davidp}.

An apparent inconsistency was noted at the level of $1/N$ corrections \cite{David}. 
Among many additional computations at this level one has to define composite operators beyond 
factorization, the simplest of which is
the operator $\alpha^2$. Below I will show that introducing the
normalization point $\mu$ -- a necessary step  not seen in \cite{David}  because
of  dimensional regularization in which the
scale separation is not explicit --  solves all would-be problems. 

\begin{figure}[t]
   \epsfxsize=4.5cm
   \centerline{\epsffile{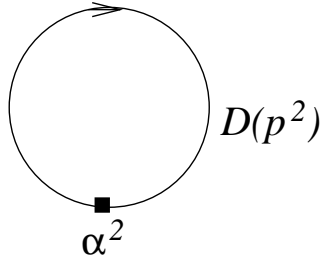}}
\caption{\small The leading contribution to $\langle [\alpha(\mu)]^2\rangle_{\rm conn}$.
The propagator of $\alpha$ is presented in (\ref{27}).
}
\label{fig4}
\end{figure}

The  vacuum expectation value of  $\alpha^2$  can be defined as follows:
\beqn
\langle [\alpha(\mu)]^2\rangle &=& m^4 + \langle [\alpha(\mu)]^2\rangle_{\rm conn} \,,\nonumber\\[2mm]
\langle [\alpha(\mu)]^2\rangle_{\rm conn} 
&=& \int _{{\rm Eucl }\,p<\mu}\, \frac{d^2p}{(2\pi)^2}\, D(p^2)\,,
\label{26}
\eeqn
where the subscript conn means the connected (nonfactorizable) part and $D(p^2)$ is the 
propagator of the $\alpha$ field known from the exact solution of the model to the leading order in $N$,
\beq
D(p^2) = - \frac{4\pi}{N}\, \frac{\sqrt{p^2(p^2+4m^2)}} 
 {\ln \frac{\sqrt{(p^2+4m^2)}+\sqrt{p^2}} 
 {\sqrt{(p^2+4m^2)}-\sqrt{p^2}}
}\,,
\label{27}
\eeq
see Fig. \ref{fig4}.

It is obvious that the connected part is suppressed
by $1/N$ compared to the factorized part.
For what follows it will be useful to rewrite the denominator  in (\ref{27}) at $p^2\gg m^2$ as
\beq
\frac{1} 
 {\ln \frac{\sqrt{(p^2+4m^2)}+\sqrt{p^2}} 
 {\sqrt{(p^2+4m^2)}-\sqrt{p^2}}
}
\equiv
\frac{\lambda(\mu )}{4\pi} \left[1 - \frac{\lambda(\mu )}{4\pi}\,\ln\frac{\mu^2}{p^2}
\right]^{-1}\left[1 + O\left(\frac{m^2}{p^2}\right)
\right].
\label{28}
\eeq
This approximation certainly becomes meaningless at $p^2\leq m^2$ since the 
expression $1 - \frac{\lambda(\mu )}{4\pi}\,\ln\frac{\mu^2}{p^2}$ vanishes at $p^2= m^2$. 

The integral in (\ref{26}) is doable and can be expressed in terms of special functions \cite{NSVZo}.
We need to examine this integral in the limit $\mu\gg m$ because only in this limit the coefficient functions
in OPE are predominantly perturbative. The result 
for $\langle [\alpha(\mu)]^2\rangle_{\rm conn}$
includes terms $O(\mu^4)$,
$O(\mu^2m^2)$, $O(m^4)$ and $O(m^6/\mu^2 )$. For simplicity will omit the latter and focus on the first three terms.
The second and the third can be established from the exact result (\ref{27}); as we will see shortly, 
they do not contain factorial divergences at all. The first term does,  see below.  We could find 
it from the exact result too. But it will be more instructive
to calculate it from the approximate formula (\ref{28}). Then the relation to renormalons will 
become more apparent. Equation (\ref{28}), being integrated from $m^2$ to $\mu^2$ (taking account of 
the remark after (\ref{28})), is perfectly sufficient to determine  	the $\mu^4$ term 
in $\langle [\alpha(\mu)]^2\rangle_{\rm conn} $. Indeed,
substituting (\ref{28}) in (\ref{27})  and keeping $m$ only in the argument of logarithms we obtain
\beqn
\langle [\alpha(\mu)]^2\rangle_{\rm conn} 
&=&
 - \frac{1}{N}\, \int _{m^2}^{\mu^2}\, p^2\, dp^2\, 
\frac{\lambda(\mu )}{4\pi} \,\sum_{k=0}^\infty \left( \frac{\lambda(\mu )}{4\pi}\,\ln\frac{\mu^2}{p^2}
\right)^{k}
\nonumber\\[4mm]
&=&
-\frac{1}{N}\, \mu^4\,\sum_{k=0}^\infty \, \left( \frac{\lambda(\mu )}{8\pi}
\right)^{k+1}\int_0^{k_*}\,dy\,y^k\, e^{-y}\,,
\label{29}
\eeqn
where 
\beq
k_* = 2 \,\ln\frac{\mu^2}{m^2}\,.
\label{30}
\eeq
If $k\leq k_*$ the integral in the right-hand side of (\ref{29}) can be extended to infinity
since the saddle point lies at $y=k$, and then this integral  produces $k!$,
\beq
\langle [\alpha(\mu)]^2\rangle_{\rm conn} 
=
-\frac{1}{N}\,\mu^4\,\sum_{k=0}^{k_*} \, \left( \frac{\lambda(\mu )}{8\pi}\right)^{k+1}\,k!\,.
\label{31}
\eeq
If $k>k_*$
the integral is saturated at $y=k_*$,  the factorial growth ceases to continue and the 
right-hand side of (\ref{29}) reduces to $O(m^4)$. 

The exact expression for $\langle [\alpha(\mu)]^2\rangle_{\rm conn}$ is well-defined and can be explicitly 
calculated. I omit a number of simple algebraic manipulations referring the reader to the original paper
 \cite{NSVZo}. The result is\footnote{I correct here a number of misprints in the expression 
 for $\langle [\alpha(\mu)]^2\rangle_{\rm conn}$ given in \cite{NSVZo}.} 
\beqn
\langle [\alpha(\mu)]^2\rangle_{\rm conn}&=& \frac{1}{N}
\left\{
 -\mu^4\left[ e^{-L} \,{\rm Ei} (L)
 \right]\right.
 \nonumber\\[3mm]
  &-&\left.  \frac{4}{L}\mu^2m^2+ 2 m^4 \left(C+\ln L -\frac{1}{L}+\frac{5}{L^2}
\right)\right\},
\label{32}
\eeqn
where
\beq
L=2\ln\frac{\mu^2}{m^2} =\left[\frac{\lambda (\mu)}{8\pi}\right]^{-1}\,,
\label{33}
\eeq
and $C\approx 0.5772...$ is the Euler constant. The $O(\mu^4)$
terms in (\ref{31}) and (\ref{32}) perfectly match each other!

Needless to say, $P(Q^2)$ does not contain the auxiliary parameter $\mu$; it depends only 
on physical parameters
$Q^2$ and $m^2$. This means that in the right-hand side of (\ref{34}) $\mu$ must cancel. 
As was mentioned, at order $N^0$ is does not appear at all. The first and the second term in (\ref{32}) appearing at the level $O(1/N)$ must be
canceled
by the corresponding contributions coming from the first and the second term in (\ref{34}). 
And they do, indeed!
The coefficient $c_0$ has a correction $\frac{\mu^4}{Q^4}\frac 1N\,\left[ e^{-L} \,{\rm Ei} (L)
 \right]$ while $c_1$ has $\lambda (\mu )\frac{\mu^2}{ Q^2}$\,. 

The terms  proportional to $m^4 $ in Eq. (\ref{32}) do not cancel. They still have a weak 
(logarithmic)
dependence on $\mu$ through $L^{-1}$ and $\ln L$. This is a manifestation of the anomalous dimension of
the operator $[\alpha(\mu)]^2$ which shows up beyond the leading (factorization) order.
 It is canceled by the corresponding logarithmic terms in $c_2(Q^2, \mu^2)$.
 
The reader interested in additional details is referred to \cite{BBK}  for a later discussion of OPE in 
a particular correlation function in the O($N)$ model at the subleasing level (i.e. O($1/N)$ 
corrections).\footnote{A remarkable feature making the model {\em different} from QCD is the fact of 
OPE convergence at 
the level of  O($N^0)$ and  O($N^{-1})$ terms. This is due the fact that at this level  
particle production thresholds do not extend to infinite energies in the O($N)$ sigma model, unlike QCD.}

Concluding this section let me mention that the O($N)$ sigma model is promising in one more
aspect: In this model at $N>3$ instantons disappear, while nothing dramatic happens to renormalons. 
Question: what replaces the instanton singularities in the Borel plane?

\section{OPE and renormalons in QCD}

After this brief digression intended to demonstrate peculiarities of
perturbation theory in strongly coupled models with the known solution let us return to QCD where no exact
solution is available.
I  will start from correlation functions of the type (\ref{currentcorr}) at large 
Euclidean $q^2$ in which OPE can be consistently 
built through separation of large- and short-distance contributions. For simplicity, for our 
illustrative purposes, I will set the separation scale at $\mu=\Lambda$ rather than at $\mu\gg\Lambda$. 
This would be inappropriate in quantitative analyses; however, my task is to reveal qualitative aspects.
For this purpose no harm will be done if I put $\mu=\Lambda$. With this convention all relevant 
expressions will dramatically simplify.

Let us have a closer look at Eqs. (\ref{renorm}) and (\ref{eight}). The unlimited factorial 
divergence in (\ref{r1}) is a direct consequence of integration over $k^2$ in (\ref{nine}) 
all the way down to $k^2=0.$ Not only this is nonsensical because
of the pole in (\ref{eight}) at $k^2=\Lambda^2$, this is {\em not} what we should  do in 
calculating coefficient functions in OPE. The coefficients must include $k^2>\Lambda^2$ by construction. The domain of small
$k^2$ (below $\Lambda^2$) must be excluded from $c_0$ and referred to the vacuum matrix 
element of the gluon operator
$G_{\mu\nu}^2$.
Indeed, in the sum in Eq. (\ref{nine}) all terms with $n>n_*$ can be written as (see Fig. \ref{fig2})
   \beqn
 \Delta D (Q^2)&=&  \frac{\alpha_s}{2}\,\sum_{n>n_*}  \left(\frac{\beta_0\alpha_s}{8\pi} \right)^n  n_*^ne^{-n_*}  
 \nonumber\\[3mm]
 &=&\frac{\alpha_s}{2}\,\sum_{n>n_*}  \frac{\Lambda^4}{Q^4}
 \label{35}
  \eeqn
  where I used the fact that $\frac{\beta_0\alpha_s(Q^2)}{8\pi}  =\frac{1}{2\ln(Q^2/\Lambda^2)}= 1/n_*$. 
  Of course, we can{\em not} calculate the gluon condensate from the above expression for the tail of the series
  (\ref{nine}) representing the large distance contribution, for a number of  reasons. In particular, 
  the value of the coefficient in front of 
  ${\Lambda^4}/{Q^4}$ remains uncertain in (\ref{35})  because Eq. (\ref{eight}) is no longer 
  valid at such momenta. 
  We do not expect the gluon Green functions used in calculation in Fig. 1 and in 
  Eq. (\ref{eight}) to retain any meaning in the nonperturbative domain of strong coupling dynamics. 
  A qualitative feature -- the power dependence 
  $(\Lambda /Q)^4$ in (\ref{35})
 -- is correct, however.
  
We note with satisfaction that the fourth 
  power of the parameter $\Lambda/Q$ which we find from this tail exactly matches the OPE contribution of
the operator  $\langle G_{\mu\nu}^2\rangle$. In Sect. \ref{illu} where we analyzed an exactly 
solvable model we could convince ourselves that this is not a coincidence. 

Summarizing this section I can say that consistent use of OPE cures the problem of the 
renormalon-related factorial divergence of the coefficients
in the $\alpha_s$ series, absorbing the IR tail of the series in the vacuum expectation value 
of the gluon operator
$G_{\mu\nu}^2$ and similar higher-order operators. Although the value of $\langle G_{\mu\nu}^2\rangle$ 
cannot be calculated from renormalons, the
very fact of its existence can be established.

\section{Sources of factorials and master formula }

From quantum mechanics we learn that the factorial divergence can arise from ``soft" fields, e.g. instantons
(see Sect. \ref{ibafd}).
In QCD the instantons are ill-defined in the IR and, strictly speaking, 
nobody knows what to do with them.\footnote{This statement is an exaggeration. 
The inquisitive reader is referred to \cite{shuryak}
for an alternative point of view on instantons in QCD vacuum.} There is a perfectly legitimate 
conceptual way out, however. If one considers QCD in the 't Hooft limit of large number of 
colors \cite{thooft}, instantons decouple. At the same time, none of the essential features of 
QCD disappears. In addition to phenomenological arguments \cite{w3}, this statement is supported
by an exact solution of a strongly coupled two-dimensional model with asymptotic freedom \cite{w4}. 

\begin{center} 
*****
\end{center} 

Now I will try to summarize the lessons we learned in a single (simplified) ``master" formula. 
At large Euclidean momenta the correlation functions of the type (\ref{currentcorr}) and similar
can be represented as
\beqn
D(Q^2)
&=&
\sum_{n=0}^{n_*^0}c_{0,n}\left(\frac{1}{\ln Q^2/\Lambda^2}\right)^n
\nonumber\\[3mm]
&+&
\sum_{n=0}^{n_*^1}c_{1,n}\left(\frac{1}{\ln Q^2/\Lambda^2}\right)^n\left(\frac{\Lambda}{Q}\right)^{d_1}
\nonumber\\[3mm]
&+&
\sum_{n=0}^{n_*^2}c_{2,n}\left(\frac{1}{\ln Q^2/\Lambda^2}\right)^n\left(\frac{\Lambda}{Q}\right)^{d_2}+ ...
\nonumber\\[3mm]
&+&
\mbox{``exponential terms"}\,.
\label{ope}
\eeqn
Equation (\ref{ope}) is simplified in a number of ways. First, it is assumed that
the currents in the left-hand side have no anomalous dimensions, and so do the operators 
appearing on the right-hand side. They are assumed to have only normal dimensions given 
by $d_i$ for the $i$-th operator. Second, I ignore the second and all higher coefficients 
in the $\beta$ function so that the running coupling is represented by a pure logarithm.
All these assumptions are not realistic in QCD.\footnote{They could be made somewhat more realistic in
${\mathcal N}=2$ super-Yang-Mills.} I stick to them to make the master formula concise. 
Inclusion of higher orders in the $\beta$ function and anomalous dimensions both on the 
left- and right-hand sides will give rise
to rather contrived additional terms and factors containing $\log\log$'s,  $\log\log\log$'s 
$(\log\log/\log)$'s, etc. 
This is a purely technical, rather than conceptual,  complication, however. 

So far I discussed the convergence of the perturbative series (explaining that the regulating parameter $\mu$
in OPE allows one to make them meaningful). The expansion (\ref{ope}) runs not only in powers
of $1/\ln Q^2$, but also in powers of $\Lambda/Q$. This is a double expansion, and the 
power series in  $\Lambda/Q$ is also infinite in its turn. Does it have a finite radius of convergence?

Needless to say, this is an important question.  The answer to it  is {\em negative}.\footnote{See also foot note 10.}
As was argued in \cite{recent,snap},
power series are factorially divergent in high orders. This is a rather straightforward
observation following  from the analytic structure of $D(Q^2)$. In a nut shell, since the cut 
in $D(Q^2)$ runs all the way to infinity along the positive real semi-axis of $q^2$, the $1/Q^2$ expansion cannot be convergent. The last line in Eq. (\ref{ope}) symbolically represents a divergent tail of the power series. 

The actual argument is somewhat more subtle than that, but the final conclusion -- that  
high-order tail of the (divergent) power series gives rise to exponentially small 
corrections (exponentially small in Euclidean, oscillating in Minkowski) -- still holds. 
The most instructive way to see it is provided by a toy model presented in Sect. 2.2 of 
\cite{snap} which refers to the 't Hooft limit. Then qualitatively one can saturate $\Pi (Q^2)$ 
by an infinite comb of equidistant infinitely narrow resonances.
For simplicity one can assume that the couplings  of these resonances to the 
current do not depend on the excitation number. Then\footnote{The factor 3 in the denominator 
of (\ref{39p}) is an approximate empiric number.}
\beq
\Pi (Q^2) = -\frac{N_c}{12\pi^2}\psi (z) +\mbox{const}\,,
\eeq
where 
\beq
z=\frac{Q^2+m_\rho^2}{3m_\rho^2}\,,
\label{39p}
\eeq
and $\psi (z) $ is Euler's $\psi$ function. 
In the Euclidean domain of positive $Q^2$
\beq
\ln z \sim -\frac{1}{2z}-\sum_{n=1}^\infty \frac{B_{2n}}{2n}\,\frac{1}{z^{2n}}\,,
\label{40}
\eeq
$B_{2n}$ stand for the Bernoulli numbers 
\beq
B_{2n} =(-1)^n \frac{2(2n)!}{(2\pi)^{2n}}\,\zeta (2n)\,,
\label{41}
\eeq
and $\zeta$ is the Riemann function. The tilde ``$\sim$" in (\ref{40}) means that the 
series in $1/Q^2$ is asymptotic: since $\zeta (2n)\sim 1$
the expansion coefficients in (\ref{40}) are obviously factorially divergent. The tail of the $1/Q^2$ series
after optimal truncation
is exponentially small. Alternatively, one can apply
the Borel procedure since the alternating signs in (\ref{41}) indicate Borel summability,
\beq
\left(\frac{1}{Q^2}\right)^n \to \frac{1}{(n-1)! }\left(\frac{1}{M^2}\right)^n\,.
\eeq
The position of the singularity in the $1/M^2$ plane is $2/(3m_\rho^2)$. 

\section{A breakthrough idea}
\label{abi}

Now I am finally ready to explain the idea first put forward in \cite{polemass,bb}. 
As was elucidated above, in the processes with OPE renormalons play no special role as 
long as the operator basis in OPE is complete, no relevant operator  is accidentally omitted. 
However, there exists a wide range of phenomena at high energies (or in heavy quark physics) 
which  do not allow one to carry out  OPE-based analyses. The most well-known example of this 
type is jet physics. Up to a certain time these processes were treated exclusively in the realm 
of perturbative QCD.
An estimate of nonperturbative effects, even as approximate as it could be, was badly needed. 
A minimalistic and urgent task was to find the power of $1/E$  (or $1/m_Q$) which controls the 
degree of fall-off of the leading nonperturbative effect. 

To this end it was suggested \cite{polemass,bb} to analyze the tails of the renormalon series. 
I hasten to add that renormalons by  no means  capture all nonperturbative effects. For instance, 
they are blind to any effects due to chiral symmetry breaking. Thus, they cannot guide us if 
chiral symmetry breaking plays a role. Hints associated 
with renormalons refer to gluons. 

The first example of the ``renormalon guidance" (that later proliferated to many other analyses)
was the so-called  heavy quark pole mass. The heavy
quark mass is a key parameter in most aspects of heavy quark physics. The pole mass was
routinely used in analyzing data. 
It is well-defined (infrared stable) and unambiguous
to any finite order in perturbation theory. This infrared stability could give an 
impression that the pole mass is well-defined in general. This misinterpretation was quite 
common in the literature in the early 1990s. 

The fact that the pole mass is {\em not} well-defined at the nonperturbative level was first noted
and emphasized
in \cite{polemass,bb}. What is even more important, a rather powerful renormalon-based tool 
was suggested for evaluating the corresponding nonperturbative contribution. The
problem arises because the pole mass is sensitive to large distance
dynamics, although this fact is not obvious  in  
perturbative calculations. Infrared contributions lead to an intrinsic
uncertainty in the pole mass of order $\Lambda $,
i.e. a  $\Lambda/m_Q$ power correction. Renormalons  produce clear evidence for this non-perturbative correction
to $m_Q^{pole}$. The signal comes from the  
factorial growth of the high order terms in the $\alpha_s$
expansion
corresponding to a  singularity residing at
$2\pi /\beta_0$ in the Borel plane.

\begin{figure}
\vspace{2.4cm}
\includegraphics{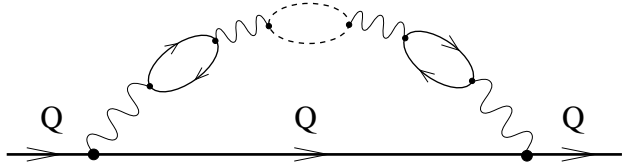}
\caption{
Perturbative diagrams leading to the IR renormalon
uncertainty in $m_Q^{\rm pole}$ of order $\Lambda$. 
The number of bubble insertions in the gluon propagator
can be arbitrary.}
\label{fig5}
\end{figure}

The renormalon contribution to the pole mass is shown in
  Fig.~\ref{fig3}. The bubble chain generates the
running of the strong coupling $\alpha_s$. To leading
order, it can be accounted for by inserting the running coupling 
constant 
$\alpha_s(k^2)$ in the integrand corresponding to the one-loop 
expression.
In  the non-relativistic regime, when   the internal momentum $|k| 
\ll 
m_Q$,   the expression is simple, 
\begin{equation}
\delta m_Q \sim - \frac{4}{3}
\int \frac{d^4k}{(2\pi )^4 i k_0} \frac{4\pi \alpha_s(-k^2)}{k^2} =
\frac{4}{3}
\int \frac{d^3\vec k}{4\pi ^2} \frac{\alpha_s (\vec k^2)}{\vec k^2}\, ,
\label{DELTAMQ}
\end{equation}
where $\alpha_s (\vec k^2)$ can be read off from Eq. (\ref{eight}) with the substitution
$Q^2\to m_Q^2$.
Expressing the running $\alpha_s(k^2)$ in terms of 
$\alpha_s(m_Q
^2)$ (note that  $\vec k^2 <m_Q^2$), and expanding in $\alpha_s(m_Q^2)$
we arrive at
\begin{equation}
\frac{\delta m_Q^{(n+1)}}{m_Q} \sim \frac{4}{3}
\frac{\alpha_s(m_Q^2)}{\pi} n!
\left( \frac{\beta_0 \alpha_s(m_Q^2)}{2\pi}\right) ^n \,.
\label{DELTAMQN}
\end{equation}
The right-hand side represents  the renormalon series. 
This series is factorially divergent and is not Borel-summable.
Moreover, in the case at hand there is no OPE  which could absorb this tail
in a higher-dimension operator. What should we do?

The question was posed and the answer given in \cite{polemass,bb}:
Following the line of reasoning applied in OPE-based processes
we should truncate the series at an optimal order and, {\em in addition},
introduce an infrared parameter $\delta m_Q$ which will absorb the renormalon tail.
Equation (\ref{DELTAMQN}) implies that
\beq
n_* \sim \ln \frac{m_Q}{\Lambda}\,,\qquad \frac{\delta m_Q}{m_Q} \sim e^{-n_*} \sim \frac{\Lambda}{m_Q}\,.
\eeq

The perturbative
expansion {\em per se} anticipates the onset of the nonperturbative 
regime (the impossibility of pinning down the 
would-be quark pole in perturbation theory to accuracy better than $\Lambda$). 

 Certainly, the 
 renormalons do not represent the dominant 
component of the infrared dynamics. 
However, they provide the ``renormalon guidance"
playing a very important role of an indicator of the presence of 
the
power-suppressed nonperturbative effects. 

\section{Renormalons  in weak  coupling problems}
\label{riwcp}

In Sect. \ref{bors} it was mentioned that the study of renormalons at weak coupling can
can shed new light on the general structure of field theory. 
According to the conjecture formulated in \cite{bo4,bo5} at weak coupling any particular 
factorially divergent contribution,
if Borel-nonsummable,  must match
a certain quasiclassical field configuration. Such configurations were identified \cite{bo5} 
in two-dimensional $CP(N-1)$ models which present a close parallel \cite{Novikov} to four-dimensional 
Yang-Mills. However, subtle  details are
not yet satisfactory. 

The instanton quarks supposedly matching the perturbative factorial divergence
 in cylindrical geometry
have action $4\pi /(N g^2)$ resulting in singularities at $4 \pi k/N$  in the Borel plane;\footnote{In four 
dimensions it is convenient to consider the $\alpha$ plane, with $\alpha\equiv g^2/4\pi$. 
In two dimensions the $g^2$ plane is more convenient.}  
here $k=1,2,..., N$ is an integer. 
If we have a look at Fig. \ref{fig3}, we will see that the positions of the renormalon singularities 
are quantized in the units of $4\pi /\beta_0$. A matching relationship can only be achieved if 
$\beta_0=$ integer $\times N$. This is the case
in $CP(N-1)$ models, but this is certainly {\em not} the case in 
Yang-Mills.\footnote{There is an observation which, perhaps, gives hope for the future. In pure Yang-Mills 
$\beta_{0}= \frac{11}{3}N$. As was noted by Khriplovich long ago \cite{khrip} (see also \cite{ucheb0}, Sect. 25.1)
the above value of $\beta_0$ has a distinct two-component structure. If one calculates $\beta_0$ in 
the physical gauge without ghosts (e.g. Coulomb), one will discover that in fact 
$\beta_{0}= 4N - \frac{1}{3}N$ where the first term
in the right-hand side presents {\em anti}screening inherent only to non-Abelian gauge theories, 
while the second term, with the fractional coefficient, is a conventional screening. } 

Let us discuss this example -- the two-dimensional $CP(N-1)$ model --  in more detail.  
Assume it is considered on a cylinder
$R_1\times S_1 ( r )$ where $r$ is the radius of the circle. At $r\to 0$ the problem reduces 
to quantum-mechanical, with the perturbative expansion being nonsummable \`a la Borel. 
The factorial divergence is not directly related to
renormalons (which are absent in quantum mechanics), but it exists. 

The corresponding singularity in the $g^2$ Borel plane
lies at $8\pi/N$. This happens to be exactly the action of two instanton quarks \cite{bo5}. 
Thus, one can (and does) achieve resurgence  through construction of the corresponding 
trans-series. What is important for the following paragraph is the fact that
action of two instanton quarks corresponds to the dimension of the lowest 
nontrivial operator in OPE, namely
$(\partial S)^2$. 

If $r\gsim \Lambda^{-1}$, we find ourselves in the strong coupling regime,
the parameter $r^{-1}$
plays the role of $\mu$, and the perturbative factorial divergence is generated by renormalons, which
should be treated, as usual, in the framework of OPE. 
Remarkably, the position of singularity in the Borel plane does {\em not} shift from $8\pi/N$.
OPE explains the renormalon  singularity at $8\pi/N$
since the first operator in OPE (after the trivial operator) has exactly the needed dimension 
and generates terms
$\exp\left(-\frac{8\pi}{N}\right)\sim \Lambda^2$. Thus, it is not ruled out that the 
positions of the leading singularity in the Borel plane is $r$-independent in the interval $r\in (0, {\rm const}\times\Lambda^{-1})$. 

This $r$ independence cannot survive  in Yang-Mills theories, because of the
$\beta_0$ factor mentioned above. What can happen in Yang-Mills, however, with luck, is a 
 smooth  $r$-dependence of the
singularity positions in the Borel plane. This question remains open. 

Cylindrical geometry exploited in \cite{bo4,bo5} 
is not the only way to make  Yang-Mills theory weakly coupled. Alternatively, one can Higgs the theory.

Assume we have   SU(2) Yang-Mills theory fully Higgsed by an expectation value of the Higgs 
doublet field, just as in the 
standard model. The theory is at weak coupling. Assume we introduce $2 n_f$ doublets of 
chiral (Weyl) fermions $(\chi_\alpha^i)^j$ and $(\psi_\alpha^i)^j$, where $\alpha$ 
and $i$ are  the Lorentz and SU(2) gauge indices, respectively, and $j=1,2,..., n_f$. 
For simplicity we will assume that
both the fermion and the Higgs  masses are the same as the mass of the $W$ bosons $M$. After 
Higgsing this theory still
has a global SU(2) symmetry. Three $W$ bosons form a triplet under this global SU(2).

This SU(2) theory has no internal anomalies. In fact, it is vector-like. With the  even 
number of doublets  
it avoids Witten's global anomaly too \cite{WGA,ucheb0}.

In this fully IR regularized theory we can repeat the analysis outlined in Sect. \ref{reno}.  
The renormalon diagram in Fig.
\ref{fig1} now yields an expression similar to that in (\ref{nine}) at $k^2\gg M^2$. However, at 
$k^2\lsim M^2$ the integral over $k^2$ in (\ref{nine}) must be cut off from below at $ M^2$ 
(with logarithmic accuracy). 

Now, in perturbation theory the large-$Q^2$ expansion of the Adler function has the form
\beqn
D(Q^2)
&=&
\sum c_{0,n}\left(\frac{1}{\ln Q^2/\Lambda^2}\right)^n + 
\sum c_{1,n}\left(\frac{1}{\ln Q^2/\Lambda^2}\right)^n\left(\frac{M}{Q}\right)^{2}
\nonumber\\[3mm]
&+&
\sum c_{2,n}\left(\frac{1}{\ln Q^2/\Lambda^2}\right)^n\left(\frac{M}{Q}\right)^{4}+ ...
\label{opeppt}
\eeqn
It is not difficult to see that (although the critical value of $n$ for the diagram in 
Fig. \ref{fig1} changes compared to that in Sect. \ref{reno})
the renormalon tail gives rise to a residual term in the second line proportional to $(\Lambda/Q)^4$, 
i.e. similar to what we have in the limit $M=0$. 

The weak coupling conjecture \cite{bo4,bo5} assumes that (\ref{opeppt})  (more exactly, the $(\Lambda/Q)^4$ term
representing its tail)
must match
a certain quasiclassical field configuration.  
At the moment the only candidate I see is the instanton-antiinstanton pair 
at a fixed (and small) separation. Is this the case? Can such a match be explicitly traced?

\section{Conclusions}

\vspace{1mm}

\noindent
1) Twenty years after its emergence \cite{polemass,bb}, the renormalon counting 
remains the only known method for evaluating nonperturbative corrections in the processes without OPE.

\vspace{1mm}

\noindent
2) Operator product expansion, with an explicit separation scale $\mu$, conceptually solves
the problem of factorial divergence of the perturbative series, at least at $N\to \infty$; 

\vspace{1mm}

\noindent
3) Factorial divergence of the $(\Lambda /Q)^k$ series  emerging in OPE at large $k$, as 
established in  \cite{recent},
 needs further explorations and an appropriate theoretical description/under\-standing.

\vspace{1mm}

\noindent
4) The resurgence program put forward in \cite{bo4,bo5} 
outlines a clear-cut parallel between factorial divergences at weak coupling on the one 
hand, which, being treated \`a la Ecalle,
result in well-defined trans-series, and the OPE-based paradigm at strong coupling, 
on the other hand. 
More thinking is required to completely understand their relationship.

\vspace{1mm}

\noindent
5) There are theories with renormalon-induced factorial divergence but no instantons (e.g. two-dimensional O($N$)
sigma model with $N>3$). 
Construction of OPE in such models is not affected by the absence of instantons. If one considers them in cylindrical geometry, at weak coupling,
finding substitutes for instanton quarks is a challenge. The first steps in this direction have been made,
but more work is needed.

\section*{Acknowledgments}

I am grateful to Martin Beneke, Ikaros Bigi, Alexei Cherman, Gerald Dunne, Sergei Monin, Mithat \"Unsal, and Arkady Vainshtein for useful comments. 

\vspace{2mm}

 This work  is supported in part by DOE grant DE-FG02- 94ER-40823. 
 

\section*{Appendix: Definitions}

We use the following convention for 
the $\beta$ function:
\begin{equation}
\label{betadef}
\beta(\alpha_s) = \mu \frac{\partial\alpha_s}{\partial \mu} = 
- \frac{1}{2\pi} \, \beta_0\,\alpha_s^2+\frac{1}{2(2\pi)^2}\beta_1\alpha_s^3+\ldots.
\end{equation}
where \cite{tlb}
\begin{equation}
\label{beta0}
\beta_0 \equiv \beta_{0\rm gluon}+\beta_{0}^{f} = \frac{11}{3}N_c -
\frac{2}{3}N_f \,.
\end{equation}
and
\beq
\beta_1 =-2\left[\frac{17}{3}N_c^2 -\frac{N_f}{6N_c}(13N^2_c-3)
\right].
\eeq
For three massless flavors $\beta_0 = 9 $.
The first two coefficients of the $\beta$ function 
are scheme-independent.


\end{document}